\documentclass[10pt,a4paper,twoside]{article}
\usepackage{epsfig}
\usepackage{array}
\usepackage{longtable}
\usepackage{baltlat6}
\usepackage{wrapfig}
\begin{document}
\ \ \vspace{-0.5mm}

\setcounter{page}{397} \vspace{-2mm}

\titlehead{Baltic Astronomy, vol.\,16, 397--408, 2007}

\vskip4mm

\titleb{A SURVEY OF COMPACT STAR CLUSTERS IN THE S-W\\
FIELD OF THE M\,31 DISK. STRUCTURAL PARAMETERS. II}

\begin{authorl}
\authorb{I.~\v{S}ablevi\v{c}i\={u}t\.{e}}{1},
\authorb{V.~Vansevi\v{c}ius}{1},
\authorb{K.~Kodaira}{2},
\authorb{D.~Narbutis}{1},
\authorb{R.~Stonkut\.{e}}{1} and
\authorb{A.~Brid\v{z}ius}{1}
\end{authorl}

\begin{addressl}
\addressb{1}{Institute of Physics, Savanori\c{u} 231,
Vilnius LT-02300, Lithuania, \\ wladas@astro.lt}
\addressb{2}{The Graduate University for Advanced Studies (SOKENDAI), \\
Shonan Village, Hayama, Kanagawa 240-0193, Japan }
\end{addressl}

\submitb{Received 2007 August 8; revised 2007 October 28; accepted 2007
October 30}

\begin{summary} The King and the EFF (Elson, Fall \& Freeman 1987)
analytical models are employed to determine the structural parameters of
star clusters using an 1-D surface brightness profile fitting method.
The structural parameters are derived and a catalogue is provided for 51
star cluster candidates from the survey of compact star clusters in the
South-West field of the M\,31 disk performed by Kodaira et al.  (2004).
\end{summary}


\resthead{Structural parameters of compact clusters in M\,31}
{I.~\v{S}ablevi\v{c}i\={u}t\.{e}, V.~Vansevi\v{c}ius, K. Kodaira et al.}

\vskip2mm
\sectionb{1}{INTRODUCTION}
\vskip1mm

Kodaira et al.  (2004; hereafter Paper~I) performed a survey of compact
star clusters in the South-West field of the M\,31 galaxy disk.  The
high-resolution Suprime-Cam imaging (Miyazaki et al. 2002) at the Subaru
Telescope (National Astronomical Observatory of Japan) enabled us to
resolve a large fraction of star clusters in M\,31.  Since the cluster
sample consists of unresolved, semi-resolved and resolved objects, one
needs to apply an appropriate method for consistent determination of
their structural parameters.

In a previous study (\v{S}ablevi\v{c}i\={u}t\.{e} et al. 2006; hereafter
Paper~II) we derived cluster structural parameters by employing the
widely used program package BAOLAB/ISHAPE (Larsen 1999).  However, the
ISHAPE algorithm is designed to work best for objects whose intrinsic
size, i.e., the Full Width at Half Maximum (FWHM) of the object
luminosity distribution, is comparable to or is smaller than the FWHM of
the star image Point Spread Function (PSF) (Larsen 2006).  Therefore, it
is difficult to apply ISHAPE for the analysis of semi-resolved clusters,
in which a few resolved stars make the fitting parameter $\chi^2$ more
sensitive to the distribution of these stars than to the general shape
of the cluster.  On the other hand, for such objects a simple direct fit
of the 1-D surface brightness profiles (see, e.g., Hill \& Zaritsky
2006) is also not appropriate due to strong alteration of the cluster
parameters by the PSF effects.  Therefore, we use a method developed to
derive the intrinsic structural parameters of clusters from 1-D surface
brightness profiles, altered by the PSF effects, using simulated star
clusters convolved with PSF.

\sectionb{2}{STAR CLUSTER SAMPLE}
\vskip1mm

The M\,31 star cluster sample studied in this paper is described in
detail in Papers~I and II.  Here we briefly remind only the most
important features of the observations.  We have used the Suprime-Cam
$V$-band frames ($5\times 2$ min exposures) of the $\sim$\,17.5\arcmin
$\times$\,28.5\arcmin size centered at $0^{\rm h}40.9\arcmin$,
$+40\degr45\arcmin$ (J2000.0).  For the study of the cluster structural
parameters we have used the $V$-band stacked mosaic image of $2
\times 3$ CCDs containing a total of $\sim$\,6K\,$\times$\,8K pixels
(pixel size $0.2\arcsec$\,$\times$\,0.2\arcsec).  The typical FWHM of
stellar images is $\sim$\,0.7\arcsec.  The morphological atlases and
photometry results for prominent compact objects ($17.5 \le V \le 19.5$)
were presented in Paper~I for 52 H$\alpha$ emission objects (KWE) and 49
non-emission clusters (KWC).

The cluster sample selected for the present study is listed in Table~1
of Paper~II.  {\it UBVRI} broad-band aperture CCD photometry data for
these clusters were published by Narbutis et al.  (2006).  In this study
we adopt the M\,31 distance modulus of $m - M = 24.5$ (e.g., Stanek \&
Garnavich 1998).

\sectionb{3}{STRUCTURAL PARAMETERS OF THE CLUSTERS}

\subsectionb{3.1}{The surface brightness profiles}
\vskip2mm

The surface brightness profiles of the clusters were derived on the
$V$-band mosaic image sub-frames of $20\arcsec\,\times\,20\arcsec$ size
using XGPHOT program implemented in IRAF (Tody 1993).  The precise
center coordinates of the clusters were determined from luminosity
weighted profiles constructed in the central regions of the sub-frames
of $4\arcsec\,\times\,4\arcsec$ size.  XGPHOT determines cluster's
ellipticity and major axis position angle based on the second order
moments of the luminosity distribution in the image.  To determine the
unbiased ellipticity and the position angle, we performed XGPHOT
photometry in circular apertures from $0.6\arcsec$ to the individual
outer radii depending on the crowding degree of the neighborhood of each
cluster.  We used apertures of increasing size with a constant step of
$0.2\arcsec$ for all clusters.  The plots of ellipticity and position
angle versus radius were used to derive the final values of both
parameters.  We determined the ellipticity and the position angle by
averaging these parameters over a ``flat'' range of radial parameter
profiles.

For the final construction of the surface brightness profiles, we
performed XGPHOT photometry in increasing elliptical apertures from
$0.1\arcsec$ out to $8\arcsec$ radius, with a step of $0.1\arcsec$ along
the major axis.  To overcome a small aperture size problem inherent to
XGPHOT, the cluster sub-frames were sub-sampled by a factor of 10 using
the IRAF {\it magnify} procedure.

The correct background estimate is essential for a reliable
representation of the cluster surface brightness profiles, especially in
their outer parts.  The sky background was analyzed interactively for
each cluster by searching for a representative region and varying its
size.  In some complex cases, additional corrections were applied,
assuming a flat cluster surface brightness profile at large radii.
\newpage

\subsectionb{3.2}{The analytical model profiles}

The structural parameters of clusters were derived by fitting
the King (1962) and the EFF (Elson et al. 1987) analytical
models. The King model is defined by the central surface brightness,
$\mu_0$, the core radius, $r_{\rm c}$, and the tidal radius, $r_{\rm
t}$:
\begin{equation}
\mu(r)=\mu_{0} \left[{\left(\displaystyle1+\displaystyle{r^{2}
\over \displaystyle r_{\rm
c}^{2}}\right)^{-{1/2}}}-{\left(\displaystyle 1+{\displaystyle
r_{\rm t}^{2} \over \displaystyle r_{\rm
c}^{2}}\right)^{-{1/2}}}\right]^2 \,. \label{King}
\end{equation}

\noindent The two following equations (Larsen 2006) were used to
compute the cluster FWHM and half-light radii, $r_{\rm h}$, from the
King profile parameters:
\begin{equation}
{\rm FWHM}=2\cdot r_{\rm c} \left[
\left(\sqrt{1/2}+{(1-\sqrt{1/2})/\sqrt{1+\left(r_{\rm t}/
r_{\rm c}\right)^2}}\right)^{-2}-1 \right]^{1/2}~,
\end{equation}

\begin{equation}
r_{\rm h} \approx 0.547\cdot r_{\rm c} \left(r_{\rm t} / r_{\rm
c}\right)^{0.486}~. \label{Kingrh}
\end{equation}

The EFF model is also defined by three parameters -- the central
surface brightness, $\mu_0$, the scale-length, $r_{\rm e}$, and  the
power-law index, $\gamma$:

\begin{equation}
\mu(r)=\mu_0 \left( 1 + {r^{2} \over r_{\rm e}^{2}}
\right)^{-{\gamma / 2}}~. \label{EFF}
\end{equation}

\noindent The three following equations (Larsen 2006) were used to
compute  FWHM and  half-light radii, $r_{\rm h}$, from
the EFF profile parameters:

\begin{equation}
{\rm FWHM}=2\cdot r_{\rm e} \sqrt{2^{2/\gamma}-1}~,
\end{equation}

\begin{equation}
r_{\rm h}=r_{\rm e}\mbox{ }\sqrt{(1/2)^{2/(2-\gamma)}-1}~.
\label{EFFrh}
\end{equation}

\noindent For the cases with $\gamma \le 2$ the total luminosity
under the profile is infinite, and, therefore, $r_{\rm h}$ is undefined.
In order to derive $r_{\rm h}$ in such extreme cases the surface
brightness profiles must be truncated at some reasonably selected
finite radii, $r_{\rm m}$:

\begin{equation}
r_{\rm h}=r_{\rm e}\mbox{ }\left[ \left[ {1 / 2} \left(
\left(1+{r_{\rm m}^2 / r_{\rm e}^2} \right)^{(2-\gamma)/2} +1
\right) \right]^{2/(2-\gamma)}-1\right]^{1/2}~. \label{EFFrh2}
\end{equation}

A discussion on the applicability and limitations of Equations
(2), (3) and (5--7) for the determination of the cluster structural
parameters is provided by Larsen (2006).

The analytical models (Equations 1 and 4) were fitted to the surface
brightness profiles via $\chi^2$ minimization.  All parameters --
$\mu_0$, $r_{\rm c}$ and $r_{\rm t}$ for the King model, and $\mu_0$,
$r_{\rm e}$ and $\gamma$ for the EFF model -- were fitted
simultaneously.  The cluster profiles were fitted from the very center
out to the radii from $1.6\arcsec$ to $4.6\arcsec$, selected considering
the cluster sizes and their neighborhood.  For the clusters KWC\,12,
KWC\,34 and KWC\,44, which do not exhibit distinct centers, the inner
fitting radius was set equal to $1.8\arcsec$, $1.2\arcsec$ and
$0.8\arcsec$, respectively.  The measured surface brightness profiles
and the fitted analytical models of six representative star clusters are
shown in Figure 1.  An averaged stellar profile is drawn in each panel.

\begin{figure}[!th]
\centerline{\psfig{figure=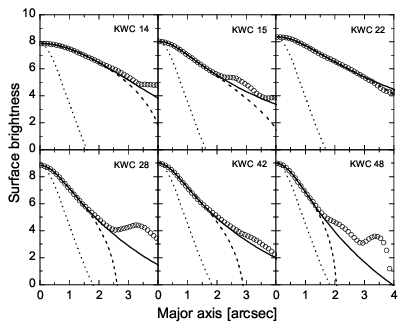,width=125mm,clip=}}
\vspace{2mm}
\captionb{1}{The surface brightness profiles of the
representative star clusters (in
the instrumental magnitude scale) are shown by open circles.
 Solid and dashed lines mark the fitted
EFF and King models, respectively; dotted lines are the averaged
stellar profiles.}
\end{figure}

\subsectionb{3.3}{Calibration grids of the cluster structural
parameters}

The cluster structural parameters derived from the 1-D surface
brightness profiles are altered by the PSF effects.  In order to
eliminate these effects, the fitted cluster parameters were calibrated
on the basis of simulated circular star clusters.  We generated
simulated cluster images with the surface brightness distribution of
King
(concentration factor ${r_{\rm t}/{r_{\rm c}}}$\,=\,1.5 -- 500; a more
conventional concentration parameter is usually defined as
$c$\,=\,log\,$({r_{\rm t}/{r_{\rm c}}})$) and EFF (${\gamma}$\,=\,1 --
40) models.  The intrinsic FWHM of these simulated clusters varied in
the range from 0.1\arcsec\ to 5.0\arcsec.  FWHM and other structural
parameters of the simulated clusters were inter-related by Equations (2)
and (5).  We convolved the simulated cluster images with the $V$-band
mosaic PSF by means of the {\it imconvol} procedure in the BAOLAB
package producing large sets of the EFF and King model based simulated
circular (ellipticity = 0.0) star clusters of different shapes and
intrinsic sizes.  PSF was constructed from well isolated stars in the
field with the DAOPHOT (Stetson 1987) program set implemented in IRAF.

\vbox{
\centerline{\psfig{figure=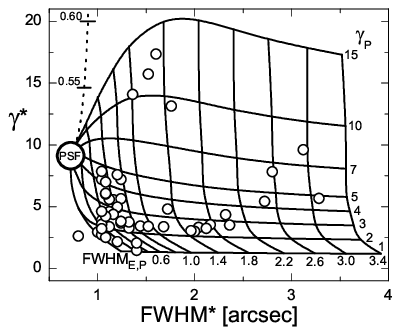,width=85mm,clip=}}
\captionb{2}{The calibration grid of the cluster
structural parameters fitted with the EFF model (FWHM$^*$,
$\gamma^*$). Open circles mark the star clusters; the large circle is
the
PSF model; the dotted line is the locus of the simulated double stars
with ticks indicating the separation between the components in FWHM
units. The numbers on the right of the grid indicate the intrinsic
power-law index, $\gamma\,_{\rm P}$, and the numbers below the
grid indicate the intrinsic FWHM$_{\rm E,P}$ (in arc-seconds) of the
simulated clusters.}
\vskip7mm
\centerline{\psfig{figure=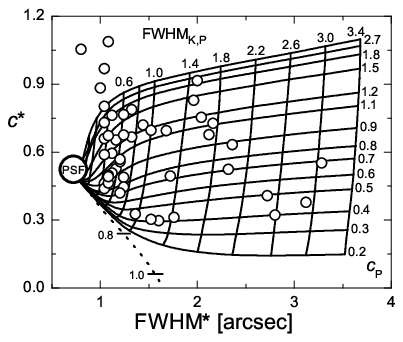,width=85mm,clip=}}
\captionb{3}{The calibration grid of the cluster
structural parameters fitted with the King model (FWHM$^*$,
$c^*$). Open circles mark the star clusters; the large circle is the PSF
model; the dotted line shows the locus of the simulated double stars
with ticks indicating the separation between the components in FWHM
units. The numbers on the right of the grid indicate the intrinsic
concentration parameter, $c\,_{\rm P}$, and the numbers on the top of
the grid indicate the intrinsic FWHM$_{\rm K,P}$ (in arc-seconds) of
the simulated clusters.}
}

The same procedure as described in section 3.1 was applied to measure
the simulated cluster images, however, all simulated clusters were
measured in circular apertures.  The clusters generated with the EFF
profile were fitted with the EFF analytical model, and those generated
with the King profile -- with the King analytical model.  The outer
fitting radius for all the EFF model simulated clusters and for the King
model simulated clusters of small intrinsic size (FWHM\,$<1.0\arcsec$)
was equal to $2.2\arcsec$.  For larger King model simulated clusters we
applied a variable outer fitting radius of $2.5\times$\,FWHM, in order
to increase the accuracy of $r_{\rm t}$ determination.

To discriminate between star clusters and simply overlapping two stars
(asterisms) we studied the 1-D profiles of simulated double stars.
Double stars were constructed from two simulated stars (generated from
PSF) of equal brightness placed at various separation distances.  A
procedure identical to that used for star clusters was applied to
measure and fit the surface brightness profiles of the simulated double
stars.  The surface brightness profiles of double stars were fitted with
the analytical models up to the outer radius of 2.2\arcsec.

The calibration grids of the cluster structural parameters (FWHM$^*$,
$\gamma^*$ for EFF; FWHM$^*$ and $c^*$ for King) fitted with analytical
models versus the intrinsic ones (FWHM$_{\rm E,P}$, $\gamma\,_{\rm P}$
for EFF; FWHM$_{\rm K,P}$ and $c\,_{\rm P}$ for King) are shown in
Figures 2 and 3. A position of PSF is indicated in the figures by a
large open circle, marked `PSF'.  The grid lines of both models converge
to this point as expected for the objects of the smallest intrinsic
size.  Dotted lines in Figures 2 and 3 show the locus of the simulated
double stars.  Note a rapid increase of fitted $\gamma^*$ and decrease
of $c^*$ with gradually increasing distance between the components of
double stars.  The star clusters and the simulated double stars fall in
distinct regions of the grids, thus allowing the recognition of some
asterisms.

The calibration grids shown in Figures 2 and 3 were employed to
determine the intrinsic (free of alteration by PSF) structural
parameters of clusters.  A bilinear interpolation was used to derive the
parameters FWHM$_{\rm E,P}$ and $\gamma\,_{\rm P}$ for the EFF models,
and FWHM$_{\rm K,P}$ and $c\,_{\rm P}$ for the King models.  The
half-light radii of clusters, $r_{\rm h}$, were computed by Eq.\,(6)
($r_{\rm h,E,P}$) and Eq.\,(3) ($r_{\rm h,K,P}$).  For the extended
clusters with $\gamma\,_{\rm P}<3$ we used Eq.\,(7), instead of
Eq.\,(6), and set the parameter $r_{\rm m}$ to be 4 times larger than
the intrinsic FWHM of the cluster, providing a lower limit for the
$r_{\rm h,E,P}$ estimate.

There are eight objects in our sample which fall outside the calibration
grid in the King model parameter space shown in Figure 3 (four of them
fall even outside the figure limits).  These objects (KWC\,02, KWC\,09,
KWC\,10, KWC\,16, KWC\,19, KWC\,24, KWE\,33 and KWE\,52) have much
larger $c^*$ values at a given FWHM$^*$ than it could be derived from
the simulated clusters.  The same eight objects (Figure 2) have the EFF
profile parameter $\gamma\,_{\rm P}<2$ (luminosity integral diverges),
suggesting very wide object wings.  Note, that the largest object
KWC\,12 falls outside the parameter range shown in Figures 2 and 3,
however, the accuracy of its structural parameters is high.

\sectionb{4}{RESULTS}

In Figure\,4 we compare the real intrinsic FWHM of star cluster derived
from the calibration grids based on the EFF (FWHM$_{\rm E,P}$) and King
(FWHM$_{\rm K,P}$) models.  Both analytical models give well correlated
FWHM -- even the largest cluster in our sample, KWC\,12, lies on the
one-to-one correspondence line.

\hbox{
\parbox[t]{58mm}{
\psfig{figure=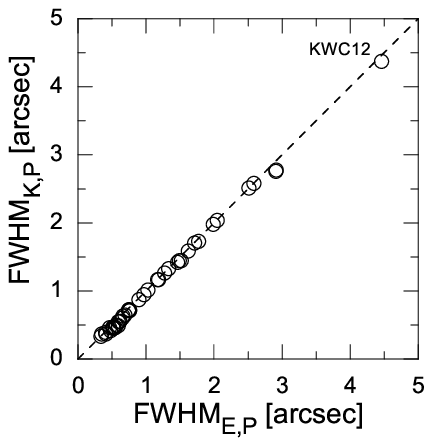,width=58mm,clip=}
\captionb{4}{The intrinsic FWHM of the clusters
derived from the calibration grids based on the EFF (FWHM$_{\rm
E,P}$) and King (FWHM$_{\rm K,P}$) models. The dashed line is the
one-to-one relation.}
}
\hskip4mm
\parbox[t]{58mm}{
\psfig{figure=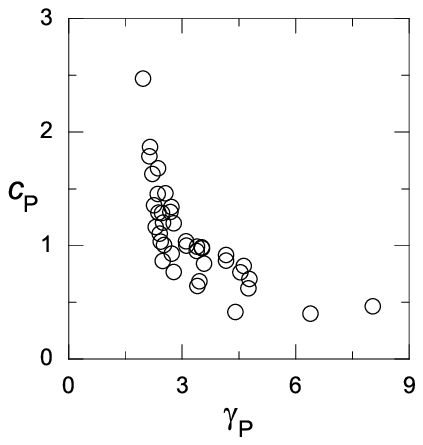,width=58mm,clip=}
\captionb{5}{The King model intrinsic concentration
parameter, $c\,_{\rm P}$, versus the EFF model intrinsic power-law
index, $\gamma\,_{\rm P}$, derived from the calibration grids.}
}
}

\begin{wrapfigure}[22]{i}[0pt]{59mm}
\psfig{figure=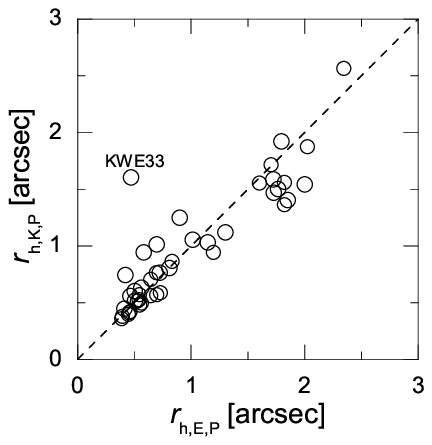,width=58mm,clip=}
\captionr{6}{Half-light radii of the star clusters
computed from Eq.\,(3) ($r_{\rm h,K,P}$ -- King) and Eqs. (6) or (7)
($r_{\rm h,E,P}$ -- EFF). The dashed line in Figures 6--10 marks the
one-to-one relation.}
\end{wrapfigure}

Figure 5 also shows a rather tight correlation of the cluster EFF model
power-law index, $\gamma\,_{\rm P}$, with the King model concentration
parameter, $c\,_{\rm P}$, suggesting robustness of the derived
structural parameters.  Comparison of the cluster intrinsic $r_{\rm h}$
determined from the EFF (Eqs.  (6)(or 7); $r_{\rm h,E,P}$) and King
(Eq.\,3; $r_{\rm h,K,P}$) model fits is shown in Figure 6.  Note the
larger (but still tolerable) scatter than in Figure 4.  The deviating
object, KWE\,33, is a young star cluster embedded in the H\,II region,
which cannot be fitted properly with the King model.

Comparisons of the cluster intrinsic parameters determined in this study
(1-D profile) with those derived with ISHAPE from Paper~II are shown in
Figures 7--10.  The largest cluster in our sample, KWC\,12, is omitted
from these figures.  Eight objects with unrealistically large $\gamma$
derived with ISHAPE ($\gamma\,_{\rm I}>7$) are not plotted.  A tight
correlation of the cluster FWHM, determined by both methods using the
EFF (FWHM$_{\rm E,P}$, FWHM$_{\rm E,I}$) and the King (FWHM$_{\rm K,P}$,
FWHM$_{\rm K,I}$) models (Figures 7 and 9), suggests that the method
used in this study produces results of comparable quality to ISHAPE.

\vbox{
\hbox{
\parbox[t]{58mm}{
\psfig{figure=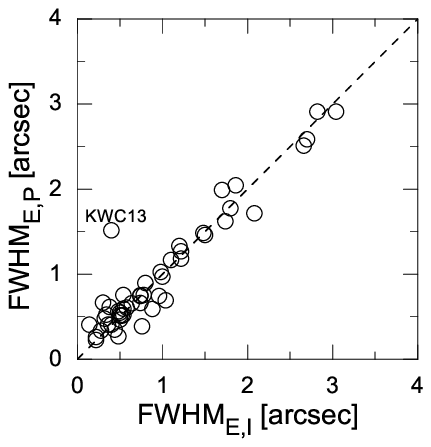,width=58mm,clip=}
\captionr{7}{Comparison of the cluster FWHM
derived by fitting the EFF models with the 1-D profile (FWHM$_{\rm
E,P}$) and ISHAPE (FWHM$_{\rm E,I}$) methods.}
}
\hskip4mm
\parbox[t]{58mm}{
\psfig{figure=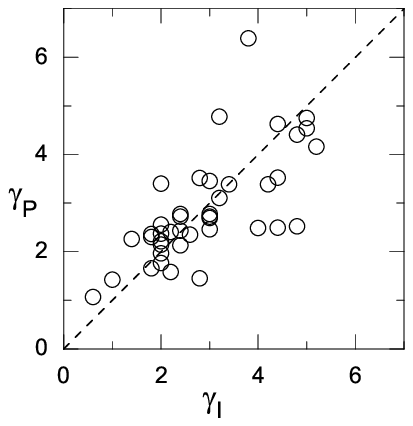,width=58mm,clip=}
\captionr{8}{Comparison of the cluster $\gamma$,
derived by fitting the EFF models with the 1-D profile
($\gamma\,_{\rm P}$) and ISHAPE ($\gamma\,_{\rm I}$) methods.}
}}
}
\vskip5mm

\vbox{
\hbox{
\parbox[t]{58mm}{
\psfig{figure=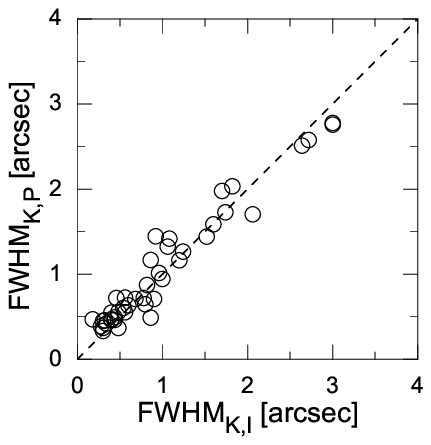,width=58mm,clip=}
\captionb{9}{Comparison of the cluster FWHM
derived by fitting the King models with the 1-D profile
(FWHM$_{\rm K,P}$) and ISHAPE (FWHM$_{\rm K,I}$) methods.}
}
\hskip4mm
\parbox[t]{58mm}{
\psfig{figure=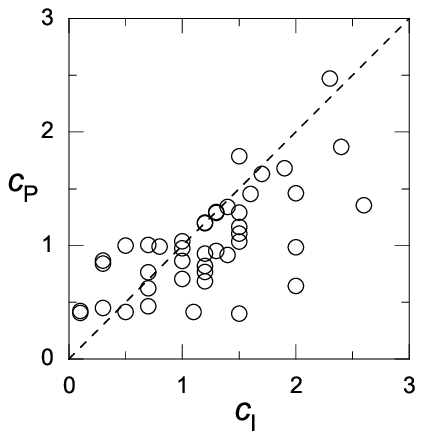,width=58mm,clip=}
\captionb{10}{Comparison of the cluster $c$
derived by fitting the King models with the 1-D profile ($c\,_{\rm
P}$) and ISHAPE ($c\,_{\rm I}$) methods.
The formal accuracy of
 the concentration parameter, $c\,_{\rm P}$, derived from 1-D profile
fitting errors, typically is in the range of $\pm$(0.1--0.2).}
}}
}

The only deviating object seen in Figure 7 is a very elongated
background galaxy candidate, KWC\,13, the displacement of which may be
due to circular simulated clusters used for constructing of the
calibration grids.  The cluster parameters, $\gamma$ ($\gamma\,_{\rm
P}$, $\gamma\,_{\rm I}$) and $c$ ($c\,_{\rm P}$, $c\,_{\rm I}$), derived
by both methods also correlate well (Figures 8 and 10).  Note that all
the objects deviating strongly from a one-to-one relation in Figures 8
and 10 cannot be fitted properly by one of the two methods (some
clusters are too much resolved for ISHAPE, and some are too small for
the 1-D profile method).  The comparison suggests that the proposed
method could effectively supplement ISHAPE, extending the workspace of
star cluster parameter determination to the domain of semi-resolved
objects.  We find that our method could be applied for the analysis of
star clusters whose intrinsic FWHM are larger than $\sim$\,50\,\% of the
FWHM of PSF.

We computed the final structural parameters of the clusters as weighted
averages of the results from the 1-D profile (this study) and the ISHAPE
(Paper~II) method (Table~1).  The parameter weights applied to different
fitting methods depended on the cluster FWHM.  For the smallest
(unresolved) clusters (FWHM\,$\leq$\,0.75\arcsec) we applied higher
weights to the ISHAPE results, and for the largest semi-resolved
clusters (FWHM\,$\geq$\,1.5\arcsec) to the results of the 1-D surface
brightness profile fits.  For the intermediate size clusters (barely
resolved) the results from both methods had comparable weights.  The
weights assigned to the fitting methods also varied, depending on the
accuracy of the fitted parameters.

In Table 1 we provide FWHM averaged from the EFF and King model
results, as well as individual values for respective analytical models
(FWHM$_{\rm E}$, $\gamma$ and FWHM$_{\rm K}$, $c$) averaged from both
methods (1-D profile and ISHAPE).  We also give averaged $r_{\rm h}$ and
individual ($r_{\rm h,E}$, $r_{\rm h,K}$) values for each analytical
model, which were computed from the intrinsic (FWHM$_{\rm E}$ and
$\gamma$; Eq.\,(6 or 7), or FWHM$_{\rm K}$ and $c$; Eq.\,(3)) parameters
and averaged from both methods.  Note that for nine clusters, which are
not fitted by the King model reliably, only half-light radii, $r_{\rm
h}$, derived from the EFF model are provided.

\begin{figure}[!th]
\centerline{\psfig{figure=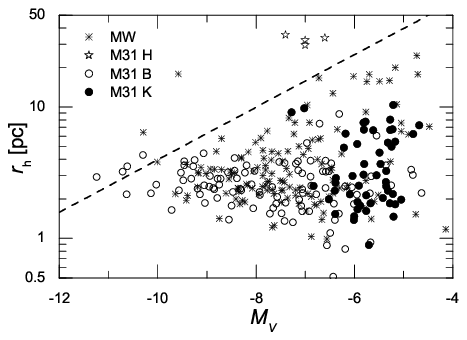,width=125mm,clip=}}
\captionb{11}{Plot of the cluster $r_{\rm h}$ versus $M_{V}$.
Filled circles mark the M\,31 clusters from our
sample; open circles -- from Barmby et al. (2002, 2007); open
stars -- the extended M\,31 clusters from Mackey et al. (2006).
The Milky Way globular clusters (Harris 1996; catalog revision:
February 2003) are shown by asterisks. The dashed line represents
the equation log\,$(r_{\rm h})=0.2\,M_V + 2.6$ (van den Bergh \&
Mackey 2004).}
\end{figure}

In the $r_{\rm h}$ vs.  $M_V$ diagram (Figure 11; van den Bergh \&
Mackey 2004), star clusters from our sample occupy the region of fainter
magnitudes than M\,31 clusters studied by Barmby, Holland \& Huchra
(2002) and Barmby et al.  (2007), and partly overlap with the parameter
region of the Milky Way globular clusters (Harris 1996).  The clusters
in our sample are brighter than $V\approx 19.5$, therefore, a cut at
$M_V\approx -5$ is due to the selection effect.  The clusters brighter
than $V\approx 17$ were saturated on Suprime-Cam images, therefore, they
were also omitted from our sample.  Note, however, that $M_V$ does not
translate directly into the cluster mass, because of a wide range of
their ages.  The clusters from the current study span a slightly wider
half-light radius ($r_{\rm h}$) range than the clusters studied by
Barmby et al.  (2002, 2007), who selected only bright and concentrated
``high probability'' globular clusters.

{\small \tabcolsep=3pt
\begin{longtable}{lcccccccc}
\multicolumn {9}{c}{\parbox{90mm}{{\normbf Table~1.}
{\norm The structural parameters of the star clusters.}}}\\
\noalign{\smallskip} \hline \noalign{\smallskip}
\multicolumn{1}{l}{Cluster}& \multicolumn{1}{c}{FWHM$\rm_E$}&
\multicolumn{1}{c}{$\gamma$}& \multicolumn{1}{c}{$r_{\rm h,E}$} &
\multicolumn{1}{c}{FWHM$\rm_K$}& \multicolumn{1}{c}{$c$} &
\multicolumn{1}{c}{$r_{\rm h,K}$} & \multicolumn{1}{c}{FWHM}&
\multicolumn{1}{c}{$r_{\rm h}$}\\
\hline \noalign{\vspace{3pt}}
\endfirsthead
\multicolumn {9}{c}{\parbox{90mm}{{\normbf Table~1.}
{\norm Continued}}}\\
\noalign{\smallskip} \hline \noalign{\smallskip}
\multicolumn{1}{l}{Cluster}& \multicolumn{1}{c}{FWHM$\rm_E$}&
\multicolumn{1}{c}{$\gamma$}& \multicolumn{1}{c}{$r_{\rm h,E}$} &
\multicolumn{1}{c}{FWHM$\rm_K$}& \multicolumn{1}{c}{$c$} &
\multicolumn{1}{c}{$r_{\rm h,K}$} & \multicolumn{1}{c}{FWHM}&
\multicolumn{1}{c}{$r_{\rm h}$}\\
\hline \noalign{\vspace{3pt}}
\endhead
\noalign{\vspace{3pt}} \hline
\endfoot
  KWC\,01  &   0.54  &  5.2   &   0.36 &  0.30 &  1.40  &   0.41 &    0.42  &  0.38   \\
  KWC\,02  &   0.24  &  1.9   &   0.48 &  0.24 &  --    &   --   &    0.24  &  0.48   \\
  KWC\,03  &   0.58  &  4.2   &   0.43 &  0.56 &  0.80  &   0.43 &    0.57  &  0.43   \\
  KWC\,04  &   0.69  &  3.1   &   0.73 &  0.70 &  0.90  &   0.58 &    0.69  &  0.66   \\
  KWC\,05  &   0.50  &  4.4   &   0.36 &  0.32 &  1.20  &   0.35 &    0.41  &  0.36   \\
  KWC\,06  &   1.30  &  $>$9  &   0.73 &  1.27 &  0.41  &   0.77 &    1.29  &  0.75   \\
  KWC\,07  &   0.54  &  3.4   &   0.49 &  0.40 &  1.30  &   0.49 &    0.47  &  0.49   \\
  KWC\,08  &   1.70  &  3.8   &   1.38 &  1.61 &  0.90  &   1.34 &    1.65  &  1.36   \\
  KWC\,09  &   0.22  &  1.8   &   0.47 &  0.34 &  --    &   --   &    0.28  &  0.47   \\
  KWC\,10  &   0.48  &  2.2   &   0.80 &  0.46 &  --    &   --   &    0.47  &  0.80   \\
  KWC\,11  &   0.59  &  4.2   &   0.44 &  0.49 &  0.87  &   0.40 &    0.54  &  0.42   \\
  KWC\,12  &   4.46  &  $>$9  &   2.52 &  4.37 &  0.30  &   2.57 &    4.42  &  2.55   \\
  KWC\,13  &   1.52  &  2.3   &   2.38 &  1.20 &  --    &   --   &    1.36  &  2.38   \\
  KWC\,14  &   2.64  &  5.2   &   1.76 &  2.60 &  0.60  &   1.73 &    2.62  &  1.74   \\
  KWC\,15  &   1.20  &  2.4   &   1.78 &  1.20 &  1.30  &   1.47 &    1.20  &  1.62   \\
  KWC\,16  &   0.41  &  1.1   &   1.42 &  1.58 &  --    &   --   &    0.41  &  1.42   \\
  KWC\,17  &   0.40  &  2.6   &   0.63 &  0.30 &  1.55  &   0.48 &    0.35  &  0.55   \\
  KWC\,18  &   0.50  &  2.4   &   0.74 &  0.44 &  1.50  &   0.66 &    0.47  &  0.70   \\
  KWC\,19  &   0.76  &  2.8   &   1.02 &  0.66 &  1.40  &   0.89 &    0.71  &  0.96   \\
  KWC\,20  &   1.46  &  $>$9  &   0.83 &  1.42 &  0.42  &   0.86 &    1.44  &  0.85   \\
  KWC\,21  &   0.75  &  2.2   &   1.24 &  0.75 &  1.67  &   1.35 &    0.75  &  1.30   \\
  KWC\,22  &   1.78  &  2.9   &   2.28 &  1.75 &  1.10  &   1.75 &    1.77  &  2.02   \\
  KWC\,23  &   1.49  &  2.3   &   2.33 &  1.50 &  1.30  &   1.83 &    1.50  &  2.08   \\
  KWC\,24  &   0.67  &  1.4   &   1.89 &  1.34 &  --    &   --   &    0.67  &  1.89   \\
  KWC\,25  &   0.75  &  3.6   &   0.64 &  0.71 &  0.84  &   0.56 &    0.73  &  0.60   \\
  KWC\,26  &   0.97  &  $>$9  &   0.55 &  0.94 &  0.46  &   0.58 &    0.95  &  0.56   \\
  KWC\,27  &   0.34  &  2.8   &   0.46 &  0.18 &  2.00  &   0.47 &    0.26  &  0.46   \\
  KWC\,28  &   0.64  &  3.2   &   0.64 &  0.54 &  1.00  &   0.49 &    0.59  &  0.57   \\
  KWC\,29  &   0.85  &  2.2   &   1.41 &  0.84 &  1.40  &   1.14 &    0.84  &  1.27   \\
  KWC\,30  &   2.51  &  4.4   &   1.82 &  2.55 &  0.60  &   1.69 &    2.53  &  1.76   \\
  KWC\,31  &   0.77  &  4.8   &   0.53 &  0.69 &  0.72  &   0.50 &    0.73  &  0.51   \\
  KWC\,32  &   2.05  &  2.8   &   2.70 &  2.00 &  1.00  &   1.82 &    2.02  &  2.26   \\
  KWC\,33  &   0.50  &  3.0   &   0.56 &  0.42 &  1.20  &   0.46 &    0.46  &  0.51   \\
  KWC\,34  &   2.91  &  8.0   &   1.71 &  2.80 &  0.50  &   1.76 &    2.86  &  1.73   \\
  KWC\,35  &   0.30  &  2.1   &   0.53 &  0.30 &  --    &   --   &    0.30  &  0.53   \\
  KWC\,36  &   1.20  &  $>$9  &   0.68 &  1.00 &  0.46  &   0.62 &    1.10  &  0.65   \\
  KWC\,37  &   0.50  &  2.1   &   0.88 &  0.50 &  1.80  &   1.04 &    0.50  &  0.96   \\
  KWC\,38  &   1.25  &  3.2   &   1.25 &  1.20 &  1.00  &   1.09 &    1.23  &  1.17   \\
  KWC\,39  &   0.74  &  4.4   &   0.54 &  0.60 &  1.20  &   0.66 &    0.67  &  0.60   \\
  KWC\,40  &   1.99  &  3.4   &   1.82 &  1.95 &  0.75  &   1.44 &    1.97  &  1.63   \\
  KWC\,41  &   1.00  &  2.3    &  1.56 &  1.00 &  1.30  &   1.22 &    1.00  &  1.39   \\
  KWC\,42  &   0.48  &  2.4    &  0.71 &  0.44 &  1.40  &   0.60 &    0.46  &  0.65   \\
  KWC\,43  &   0.40  &  2.3    &  0.63 &  0.36 &  1.80  &   0.75 &    0.38  &  0.69   \\
  KWC\,44  &   2.91  &  4.7    &  2.03 &  2.80 &  0.65  &   1.92 &    2.86  &  1.98   \\
  KWC\,45  &   0.54  &  3.2    &  0.54 &  0.46 &  1.00  &   0.42 &    0.50  &  0.48   \\
  KWC\,46  &   2.00  &  3.5    &  1.77 &  2.00 &  0.95  &   1.74 &    2.00  &  1.75   \\
  KWC\,47  &   0.74  &  3.0    &  0.84 &  0.60 &  1.30  &   0.73 &    0.67  &  0.78   \\
  KWC\,48  &   0.52  &  4.4    &  0.38 &  0.42 &  1.00  &   0.38 &    0.47  &  0.38   \\
  KWC\,49  &   0.36  &  3.0    &  0.41 &  0.32 &  1.30  &   0.39 &    0.34  &  0.40   \\
  KWE\,33  &   0.44  &  2.1    &  0.78 &  0.48 &  --    &   --   &    0.46  &  0.78   \\
  KWE\,52  &   0.14  &  2.2    &  0.23 &  0.10 &  --    &   --   &    0.12  &  0.23   \\
\end{longtable}}
{\bf Note:} FWHM -- the averaged final FWHM; FWHM$\rm_E$ and
FWHM$\rm_K$ -- FWHM derived from the EFF and King models,
respectively (in arc-seconds); $\gamma$ -- the EFF model power-law
index; $c$ -- the King model concentration parameter,
$c$\,=\,log\,$(r_{\rm t}/r_{\rm c})$; $r_{\rm h}$ -- the averaged final
half-light radius; $r_{\rm h,E}$ and $r_{\rm h,K}$ -- the half-light
radii derived from the EFF and King models, respectively (in
arc-seconds). The final FWHM and $r_{\rm h}$ (presented in the
last two columns) are derived by averaging FWHM$\rm_E$ and
FWHM$\rm_K$; $r_{\rm h,E}$ and $r_{\rm h,K}$, respectively.

\sectionb{5}{SUMMARY}

We apply the 1-D surface brightness profile fitting method to determine
structural parameters of star clusters by fitting the King and the EFF
models and demonstrate its performance on 51 cluster candidates from the
survey of compact star clusters in the M\,31 disk (Paper~I).  The
structural parameters of clusters are derived from the $V$-band images
using the calibration grids based on simulated star clusters.

A set of simulated clusters with predefined structural parameters was
generated and convolved with the $V$-band PSF.  The surface brightness
profiles of the simulated clusters were derived from the aperture
photometry data, and their structural parameters were determined by
fitting the analytical EFF and King models in the same manner as for the
real star clusters.  The structural parameter calibration grids were
constructed on the basis of the simulated cluster data.

We find that the 1-D surface brightness profile fitting method works
well for the majority of clusters in our sample.  The derived intrinsic
FWHM and half-light radii, $r_{\rm h}$, are virtually independent on the
analytical models employed.  The structural parameters determined in
this study (1-D profile fit) are in good agreement with those derived
for the same star clusters by ISHAPE (Paper~II).  Therefore we conclude
that the 1-D profile fitting method is a robust tool for the
determination of structural parameters of star clusters, and may be
effectively applied for the analysis of clusters whose intrinsic FWHM
are larger than $\sim$\,50\,\% of the FWHM of PSF.

\thanks{We are thankful to our referee S. Larsen for his constructive
suggestions.  This work was financially supported in part by a Grant of
the Lithuanian State Science and Studies Foundation.  The research has
made use of SAOImage DS9, developed by the Smithsonian Astrophysical
Observatory.}

\References

\refb Barmby~P., Holland~S., Huchra~J. 2002, AJ, 123, 1937

\refb Barmby~P., McLaughlin~D.~E., Harris~W.~E., Harris~G.\,L.\,H.,
Forbes~D.~A. 2007, AJ, 133, 2764

\refb Elson~R.\,A.\,W., Fall~S.~M., Freeman~K.~C. 1987, ApJ, 323, 54

\refb Harris~W.~E. 1996, AJ, 112, 1487

\refb Hill~A., Zaritsky~D. 2006, AJ, 131, 414

\refb King~I. 1962, AJ, 67, 471

\refb Kodaira~K., Vansevi\v{c}ius~V., Brid\v{z}ius~A., Komiyama~Y.,
Miyazaki~S., Stonkut\.{e}~R., \v{S}ablevi\v{c}i\={u}t\.{e}~I.,
Narbutis~D. 2004, PASJ, 56, 1025 (Paper~I)

\refb Larsen~S.~S. 1999, A\&AS, 139, 393

\refb Larsen~S.~S. 2006, {\it An ISHAPE User's Guide} (ver. Oct. 23,
2006; available at http://www.astro.uu.nl/$\sim$larsen/baolab/),
p. 14

\refb Mackey~A.~D., Huxor A., Ferguson A.\,M.\,N. et al. 2006, ApJ, 653,
L105

\refb Miyazaki~S., Komiyama Y., Sekiguchi M. et al. 2002, PASJ, 54, 833

\refb Narbutis~D., Vansevi\v{c}ius~V., Kodaira~K.,
\v{S}ablevi\v{c}i\={u}t\.{e}~I., Stonkut\.{e}~R., Brid\v{z}ius~A.
2006, Baltic Astronomy, 15, 461

\refb \v{S}ablevi\v{c}i\={u}t\.{e}~I.,Vansevi\v{c}ius~V.,
Kodaira~K., Narbutis~D., Stonkut\.{e}~R., Brid\v{z}ius~A. 2006,
Baltic Astronomy, 15, 547 (Paper~II)

\refb Stanek~K.~Z., Garnavich~P.~M. 1998, ApJ, 503, L131

\refb Stetson~P.~B. 1987, PASP, 99, 191

\refb Tody~D. 1993, in {\it Astronomical Data Analysis Software
and Systems II}, eds. R.~J.~Hanisch, R.~J.~B.~Brissended,
J.~Barnes, ASP Conf. Ser. 52, ASP, San Francisco, 173

\refb van den Bergh~S., Mackey~A.~D. 2004, MNRAS, 354, 713

\end{document}